\documentclass{rcdj}

\begin{document}

\setcounter{page}{161}
\journal{REGULAR AND CHAOTIC DYNAMICS, V.\,7, \No2, 2002}
\title{ON THE INTEGRATION THEORY\\
OF EQUATIONS OF NONHOLONOMIC
MECHANICS\thanks{Advances in mechanics USSR, V.~8, \No3, pp.~85--107, 1985.}}
\runningtitle{ON THE INTEGRATION THEORY OF EQUATIONS OF NONHOLONOMIC
MECHANICS}
\runningauthor{V.\,V.\,KOZLOV}
\authors{V.\,V.\,KOZLOV}
{Department of Mechanics and Mathematics\\
Moscow State University,
Vorob'ievy Gory\\
119899, Moscow, Russia}
\amsmsc{37J60, 37J35}
\doi{10.1070/RD2002v007n02ABEH000203}
\abstract{The paper deals with the problem of integration of equations of motion in
nonholonomic systems. By means of well-known theory of the differential
equations with an invariant measure the new integrable systems are
discovered. Among them there are the generalization of Chaplygin's
problem of rolling nonsymmetric ball in the plane and the Suslov problem
of rotation of rigid body with a fixed point. The structure of dynamics of
systems on the invariant manifold in the integrable problems is shown.
Some new ideas in the theory of integration of the equations in
nonholonomic mechanics are suggested. The first of them consists in using
known integrals as the constraints. The second is the use of resolvable
groups of symmetries in nonholonomic systems. The existence conditions of
invariant measure with analytical density for the differential equations
of nonholonomic mechanics is given.}

\maketitle

\section{Introduction}

The integration theory of equations of motion for mechanical systems with
nonholonomic constraints isn't so complete as in the case of systems with
holonomic constraints. This fact has many reasons. First, the equations of
nonholonomic mechanics have a more complex structure than the Lagrange
equations, which describe the dynamics of systems with integrable
constraints. For example, a nonholonomic system can't be characterized by
the only function of its state and time (cf.~[1], ch.~XXIV). Second, the
equations of nonholonomic mechanics have no invariant measure in the
general case (a~simple example is given in section~5). The point is that
nonholonomic constraints may be realized by action of complementary forces
of viscous anisotropic friction with a large viscosity coefficient~([3]).
The absence of an invariant measure is a characteristic property of
systems with friction. In limit, the anisotropic friction is compatible
with the conservation of total energy. But asymptotically stable
equilibriums or limit cycles may arise on the manifolds of energy levels
(cf.~[4]), and this is the reason for nonexistence of additional
``regular''
integrals of motion.

The most popular method to integrate the equations of nonholonomic
dynamics is based on the use of the available first integrals or the
``conservation laws'': if a Lie group that acts on a position space
preserves the Lagrangian and if vector fields that generate this group are
the fields of possible velocities then the equations of motion have the
first ``vector'' integral or the generalized integral of kinetic moment
[5,6]. A number of problems of nonholonomic dynamics was solved by this
method, among them, we note especially Chaplygin's problem on an
asymmetrical ball rolling over a horizontal plane [5].

Attempts to generalize the Hamilton\f Jacobi method to the systems with
nonholonomic constraints were non-effective as well as attempts to present
the equations of nonholonomic dynamics in the form of Hamiltonian
canonical equations. It turned out that with the help of the
Hamilton\f Jacobi generalized method it is possible to find at most only
some special solutions of the motion equations. This paper contains the
detailed analysis of these questions.

Another general approach to the integration of nonholonomic equations is
based on the theory of Chaplygin's reducing factor ([5]): one try to
obtain a change of time (different along different trajectories), such
that the equations of motion are presented as Lagrange or Hamilton
equations. Though such change exists in exceptional cases only, it allows
to solve a number of new problems of nonholonomic dynamics (cf. [5]). Let
us note that the equations of motion sometimes may be reduced to the
Hamiltonian form by other reasoning (see section 5).

The list of exactly solvable problems of nonholonomic dynamics isn't long:
almost complete information may be found in the books [1,5,8]. In this
work we present some new integrable problems, consider the characteristic
features of behavior of nonholonomic systems' trajectories in the phase
space, and propose some general theoretical reasonings on methods of
integrating the equations of nonholonomic dynamics.

\section{Differential equations with an integrable measure}


Let us consider a differential equation
\begin{equation}
\label{2.1}
\dot{x}=f(x), \q x\in\mR^n
\end{equation}
and let $g^t$ be its phase flow. Suppose (\ref{2.1}) has an integral
invariant with some smooth density~$M(x)$, i.e. for any measurable
domain~$D\sbs R^n$ the following equation holds for all~$t$
\begin{equation}
\label{2.2}
\intl_{g^t(D)}M(x)\,dx=\intl_D M(x)\,dx.
\end{equation}

Let us recall the well-known Liouville assertion: a smooth
function~$f\colon\mR^n\to\mR$ is a density of an invariant \eq*{ \int
M(x)\,dx } \phantom{P} if and only if $\div(Mf)\equiv 0$. If $M(x)>0$ for
all $x$ then~(\ref{2.2}) defines a measure in~$\mR^n$ invariant with
respect to the action of~$f$. The existence of an invariant measure
simplifies the integration of a differential equation; for example, in the
case of~$n=2$ the equation is always integrable in quadratures.
According to Euler,~$M$ is also referred to as an integrating factor.

\begin{teo}(1)
Suppose system \eqref{2.1} with invariant measure~\eqref{2.2} has~$n-2$
first integrals $F_1\dts F_{n-2}$. Suppose the functions $F_1\dts F_{n-2}$ are
independent on the invariant set ${\t E}_c=\bigl\{x\in\mR^n\colon F_s(x)=c_s$,
$1\le s\le n-2\bigr\}$.

Then

$1)$ the solutions of \eqref{2.1} that belong to ${\t E}_c$ may be found by
quadratures. If~$L_c$ is a connected compact component of the level set~${\t E}_c$
and~${f\ne 0}$ on~${\t L}_c$ then

$2)$~${\t L}_c$ is a smooth surface diffeomorphic to a two-dimensional
torus,

$3)$~it is possible to choose angle variables ${x,\,y\mod{2\pi}}$ on~${\t L}_c$
so that, after the change of variables, system~\eqref{2.1} on~${\t L}_c$ would
have the following form
\begin{equation}
\label{2.3}
\dot{x}=\frac{\lm}{\Fi(x,\,y)}, \q \dot{y}=\frac{\mu}{\Fi(x,\,y)},
\end{equation}
where $\lm,\,\mu=\const$, $|\lm|+|\mu|\ne 0$
and~$\Fi$ is a smooth positive function $2\pi$\1periodical with
respect to~$x$ and~$y$.
\end{teo}

Let us mention the main points of the proof. Since the vector field~$f$
is tangent to~${\t E}_c$,  differential equation~(\ref{2.1}) is bounded
on~${\t E}_c$. This equation on ${\t E}_c$ has an integral invariant
\eq*{
\int\frac{M\,d\sg}{V_{n-2}},
}
where $d\sg$ is the element of area of ${\t E}_c$ considered as a surface
embedded into~$\mR^n$, $V_{n-2}$ is the ${(n-2)}$\1dimentional volume of the
parallelepiped in~$\mR^n$, the gradients of $F_1\dts F_{n-2}$ being its sides.
Now the integrability by quadratures on~$E_c$
follows from Euler's remark. The first conclusion of theorem~1 (which was
firstly mentioned by Jacobi) is proved by this reasoning. The second
conclusion is the well-known topological fact: any connected, compact,
orientable,
two-dimensional manifold that admits a tangent field without singular
points is diffeomorphic to a two-dimensional torus. The third conclusion is,
in fact, the Kolmogorov theorem on reduction of differential
equations on a torus with a smooth invariant measure [9].

Equations (\ref{2.3}) have invariant measure
\eq*{
\iint\bigl|\Fi(x,\,y)\bigr|\,dx\wedge dy.
}
By averaging the right-hand sides of (\ref{2.3}) with respect to this measure
we get the differential equations
\begin{equation}
\label{2.4}
\dot{u}=\frac{\lm}{\nu}, \q \dot{v}=\frac{\mu}{\nu}; \q
\nu=\frac{1}{4\pi^2}\intl_0^{2\pi}\intl_0^{2\pi}\Fi\,dx\,dy.
\end{equation}

\begin{pro}(1)
Let $\Fi\colon T^2\to\mR$ be a smooth {\rm(}analytical\/{\rm)} function. Then
for almost all pairs $(\lm,\,\mu)\in\mR$ there exists a smooth
{\rm(}analytical\/{\rm)}
change of angle variables $x,\,y\to u,\,v$ that reduces~\eqref{2.3}
to~\eqref{2.4}.
\end{pro}

The proof is presented in [9,10]. Note that if (\ref{2.3}) can(not) be
reduced to~(\ref{2.4}) for a pair~$(\lm,\,\mu)$ then the same is true for
all pairs $(\vk\lm,\,\vk\mu)$, $\vk\ne 0$. So, the property of reducibility
depends on arithmetical properties of~$\frac{\lm}{\mu}$ that is referred
to as the rotation number of the tangent vector field on
$T^2=\bigl\{x,\,y\mod{2\pi}\bigr\}$.

\begin{pro}(2)
Let $\Fi(x,\,y)=\sum\vfi_{m,n}\exp i(mx+ny)$,
$\vfi_{m,n}=\ol{\vfi}_{-m,-n}$.
If~\eqref{2.3} may be reduced to~\eqref{2.4} by a differentiable change of
angle variables $u=u(x,\,y)$, $v=v(x,\,y)$ then
\begin{equation}
\label{2.5}
\suml_{|m|+|n|\ne 0}\Bigl|\frac{\vfi_{m,n}}{m\lm+n\mu}\Bigr|^2<\fy.
\end{equation}
\end{pro}

If the ratio $\frac{\lm}{\mu}$ is rational then the torus $T^2$ is stratified into
a family of closed trajectories. In this case the reducibility
condition is equivalent to the equality of periods of rotation
for different closed trajectories.

In the general case (the Fourier decomposition of $\Fi$ contains harmonics)
the points $(\lm,\,\mu)\in\mR^n$ with rationally independent~$(\lm,\,\mu)$,
for which series~(\ref{2.5}) diverges, are everywhere dense in~$\mR^n$.
The questions of reducibility of~(\ref{2.3}) are discussed in [9]. The
general properties of solutions of~(\ref{2.3}) see  in [10].

\section{S.\,A.\,Chaplygin's problem}


Let us consider as an example the problem of
rolling of a balanced, dynamically
non-symmetric ball on a horizontal rough plane (see [5]).
The motion of the ball is described by the following system of
equations in $\mR^6=\mR^3\{\om\}\x\mR^3\{\gam\}$:
\begin{equation}
\label{3.1}
\begin{gathered}
\dot{k}+\om\x k=0, \q \dot{\gam}+\om\x\gam=0;\\
k=I\om+ma^2\gam\x(\om\x\gam).
\end{gathered}
\end{equation}

Let $\om$ be the vector of the angular rotation velocity of the ball,~$\gam$
the unit vertical vector,~$I$ the tensor of inertia of the
ball with respect to its center,~$m$ the mass of the ball, and~$a$ its radius.
These equations have the invariant measure with density
\begin{equation}
\label{3.2}
M=\frac{1}{\sqrt{(ma^2)^{-1}-\<\gam,\,(I+ma^2 {\t E})^{-1}\gam\>}}, \q
{\t E}=\|\dl_{ij}\|.
\end{equation}

Taking into account the existence of four independent integrals
$F_1=\<k,\,\om\>$, $F_2=\<k,\,\gam\>$, $F_3=\<\gam,\,\gam\>$,
$F_4=\<k,\,k\>$,
we see that~(\ref{3.1}) is integrated by quadratures. Note that system of
equations~(\ref{3.1}) has no equilibriums on the non-critical level
sets~${\t E}_c$. Indeed, if~$\dot{\gam=0}$, then~$\om$ and~$\gam$ are
linearly dependent. This fact implies the linear dependence
of~$dF_1$ and~$dF_2$. The simplest case of integration by quadratures of
equations~(\ref{3.1}) is the case of zero value of the constant in the
``area''
integral~$F_2$. In elliptic coordinates~$\xi,\,\eta$ on the Poisson
sphere~$\<\gam,\,\gam\>=1$ the equations of motion on the level~${\t E}_c$
are reduced to the following form
$$
\begin{gathered}
\dot{\xi}=\frac{\sqrt{P_5(\xi)}}{\xi(\xi^{-1}-\eta^{-1})\,\Fi(\xi,\,\eta)};
\q
\dot{\eta}=\frac{\sqrt{P_5(\eta)}}{\eta(\xi^{-1}-\eta^{-1})\,\Fi(\xi,\,\eta)};\\
\Fi=\sqrt{(a-\xi)(a-\eta)}.
\end{gathered}
$$

Coefficients of the polynomial $P_5$ of the fifth order and the constant~$a$
depend on parameters of the problem and on the constants of the first
integrals (see [5] for details).
The variables~$\xi,\,\eta$ ranges over different closed intervals
$a_1\le\xi\le a_2$, $b_1\le\eta\le b_2$
where~$P_5$ is nonnegative. The uniformizing substitution
\begin{equation}
\label{3.3}
\begin{aligned}
x & =\lm\intl_{a_1}^\xi\frac{z\,dz}{\sqrt{P_5(z)}}, & \q
\lm^{-1} & =\frac{1}{\pi}\intl_{a_1}^{a_2}\frac{z\,dz}{\sqrt{P_5(z)}},\\
y & =\mu\intl_{b_1}^\eta\frac{z\,dz}{\sqrt{P_5(z)}}, &
\mu^{-1} & =\frac{1}{\pi}\intl_{b_1}^{b_2}\frac{z\,dz}{\sqrt{P_5(z)}},
\end{aligned}
\end{equation}
introduces the angle variables $x,\,y\mod{2\pi}$ on ${\t E}_c$, and the
equations of motion take form~(\ref{2.3})
\begin{equation}
\label{3.4}
\begin{gathered}
\dot{x}=\frac{\lm}{\Fi(x,\,y)}, \q \dot{y}=\frac{\mu}{\Fi(x,\,y)},\\
\Fi=\bigl(\xi^{-1}(x)-\eta^{-1}(y)\bigr)\sqrt{\bigl(a-\xi(x)\bigr)\bigl(a-\eta(y)\bigr)}.
\end{gathered}
\end{equation}
Here $\xi(x)$ and $\eta(y)$ are $2\pi$\1periodic functions of~$x$ and~$y$
arising as the inversions of Abelian integrals~(\ref{3.3}).

These equations imply

\begin{pro}(3)
The rotation number of a tangent vector field on
two-dimensional invariant tori in Chaplygin's problem is equal
to the ratio of real periods of the Abelian integral
\eq*{
\int\frac{z\,dz}{\sqrt{P_5(z)}}.
}
\end{pro}

\begin{rem*}
This assertion is true for integrable problems of
dynamics of a heavy body with a fixed point defined by
the system of Euler-Poisson equations (see~[10]). Since the Euler-Poisson
equations are Hamiltonian, by the Liouville theorem, in integrable cases
they always can be reduced to form~(\ref{2.4})
on two-dimensional invariant tori. It seems that equations~(\ref{3.4})
have no such property; inequality~(\ref{2.5}) is not fulfilled
on all invariant non-resonant tori.
\end{rem*}

Let us make a change of time $t\to\tau$ by the formula
\begin{equation}
\label{3.5}
dt=\sqrt{(a-\xi)(a-\eta)}\,d\tau.
\end{equation}
Equations (\ref{3.4}) preserve their form but the variables $x,\,y$ in
function~$\Fi$ are separated
\eq*{
\Fi=\xi^{-1}(x)-\eta^{-1}(y).
}

\begin{pro}(4)
Suppose that $\mu$ and $\lm$ in \eqref{2.3} are nonzero and
\eq*{
\Fi=\Fi'(x)+\Fi''(y).
}
Then equations \eqref{2.3} can be reduced to form \eqref{2.4} by an
invertible change of angle variables on~$T^2$.
\end{pro}

The proof is presented in [10]. Note that if $\Fi=\Fi'(x)+\Fi''(y)$,
then series~(\ref{2.5})
\eq*{
\sum_{n\ne
0}\Bigl|\frac{{\vfi'}_n}{n\lm}\Bigr|^2+\Bigl|\frac{{\vfi''}_n}{n\mu}\Bigr|^2
}
converges for all $\lm,\,\mu\ne 0$.

So, taking into account change of time (\ref{3.5}) one may reduce~(\ref{3.4})
to the form
\begin{equation}
\label{3.6}
\frac{du}{d\tau}=U, \q \frac{dv}{d\tau}=V,
\end{equation}
where $U$ and $V$ depend on the constants of the first integrals only,
and~$U,\,V\ne 0$.
This result can cause the temptation to use Chaplygin's reducing multiplier
theorem: if we can reduce~(\ref{3.1}) using change of time~(\ref{3.5}) to
the Euler-Lagrange equations of some variational problem (they are written,
as well as the classical Euler-Poisson equations, in redundant variables)
then according to the Liouville theorem, the equations of motion on the
two-dimensional invariant tori in some angle variables~$u,\,v\mod{2\pi}$
have form~(\ref{3.6}). It is possible to show that this method does not
lead to the goal. In conclusion, note that S.\,A.\,Chaplygin himself never
considered the problem of the ball's rolling in connection with the reducing
multiplier theory.

\section{A generalization of S.A.Chaplygin's problem}


We are going to show that the problem of rolling of a balanced, dynamically
non-symmetric ball on a rough plane is still integrable
(in the sense of section~1), if the particles of the ball are attracted
by this plane proportionally to the distance. Since the center of mass
of the ball coincides with its geometrical center, we can calculate
the potential by the formula
\begin{equation}
\label{4.1}
V(\gam)=\frac{\eps}{2}\int\<{\br},\,{\bs \gam}\>^2 \,
dm=\frac{\eps}{2}\<{\bI\gam},\,\gam\>,
\end{equation}
where $\gam$ is the unit vertical vector, ${\br}$ is the radius vector of
particles of the ball, ${\bI}$ is the tensor of inertia of the ball with
respect to its center. The attraction forces generate the
rotational moment
\eq*{
-\int{\br}\x\bigl(\eps\<{\br},\,\gam\>\gam\bigr)\,dm=-\eps\int\<{\br},
\,\gam\>({\br}\x\gam)\,dm=\gam\x {V'}_\gam=\eps\gam\x{\bI}\gam.
}
In order to obtain the moment of forces with respect to the contact
point, it is necessary to add the moment of the combined force
\eq*{
\eps\int\<{\br},\,\gam\>\gam\,dm=\eps\Bigl\<\int{\br}\,dm,\,\gam\Bigr\>\gam,
}
which is equal to zero, since the center of mass of the ball coincides
with its geometrical center.

According to the theorem about the change of kinetic moment behavior
with respect to
the contact point (see [5], [6]), the equations of rolling of the ball
can be presented in the following form
\begin{equation}
\label{4.2.}
\dot{k}+\om\x k=\eps\gam\x{\bI}\gam, \q \dot{\gam}+\om\x\gam=0.
\end{equation}

\begin{teo}(2)
Differential equations \eqref{4.2.} are integrable by quadratures.
\end{teo}

Indeed, they have four independent integrals
\begin{equation*}
\begin{gathered}
F_1=\<k,\,\om\>+\eps\<{\bI}\gam,\,\gam\>, \q F_2=\<k,\,\gam\>, \q
F_3=\<\gam,\,\gam\>=1,\\
F_4=\<k,\,k\>-\<A\gam,\,\gam\>,
\end{gathered}
\end{equation*}
where elements $A_i$ of a diagonal matrix $A$ are expressed through
the principal moments of inertia~$I_i$ by the formulae
\eq*{
A_1=\eps(I_2+ma^2)(I_3+ma^2), \ldots
}
Since equations (\ref{4.2.}) have the invariant measure with density~(\ref{3.2}),
they are integrable by theorem~1. It would be interesting to
integrate this equation explicitly and test if proposition 3 remains true
for equations~(\ref{4.2.}).

Note that the problem of rotation of a body about a fixed point
in an axisymmetric force field with potential~(\ref{4.1}) is also
integrable~([1]). In addition to the classical integrals~$F_1$, $F_2$, $F_3$,
there is the integral~$F_4$, where one must put~${\ba=0}$. This integral was
found independently
by Clebsh in the problem on motion of a body in an ideal fluid and
by Tisseran, who investigated rotational motion of heavenly bodies.

\section{G.\,K.\,Suslov's problem and its generalization}

Following G.\,K.\,Suslov ([11], ch. 53), we consider the problem of
rotation about a fixed point of a body with the nonintegrable
constraint~$\<{\ba},
\,\om\>=0$, where~${\bf a}$ is a vector that is constant in the moving frame
of reference.
Suppose that the body rotates in an axisymmetric force field with the
potential~$V(\gam)$. Following the method of Lagrange multipliers,
we write down the equations of motion ([11], ch.~46):
\begin{equation}
\label{5.1}
{\bI}\dot{\om}+\om\x{\bI}\om=\gam\x V'_\gam+\lm{\ba}, \q
\dot{\gam}+\om\x\gam=0, \q \<{\ba},\,\om\>=0.
\end{equation}

Using the constraint equation $\<{\ba},\,\om\>=0$, the Lagrange factor can
be expressed as the function of~$\om$ and~$\gam$
\eq*{
\lm=-\bigl\<{\ba},\,I^{-1}({\bI}\om\x\om)+I^{-1}
(\gam\x V''_\gam)\bigr\>\bigl/\<{\ba},\,I^{-1}{\ba}\>.
}

Equations (\ref{5.1}) always have three independent integrals:
\begin{equation*}
\begin{gathered}
F_1=\<{\bI}\om,\,\om\>\bigl/2+V(\gam), \q F_2=\<\gam,\,\gam\>, \q
F_3=\<\ba,\,\om\>.
\end{gathered}
\end{equation*}

For real motions, $F_2=1$, $F_3=0$. In this case, we can reduce the
problem of integration
of equations~(\ref{5.1}) to the problem of existence of an invariant measure
(the existence isn't evident) and the fourth independent integral.

\begin{pro}(5)
If $\ba$ is an eigenvector of operator $\bI$, then the phase flow of
system~\eqref{5.1}
preserves the ``standard'' measure in $\mR^6=\mR^3\{\om\}\x \mR^3\{\gam\}$.
\end{pro}

To prove the proposition we have to verify the following fact: the
divergence of the right-hand side of~(\ref{5.1}) is equal to zero
as~$\bI\ba=\mu \ba$.

\wfig<bb=0 0 35.6mm 32.3mm>{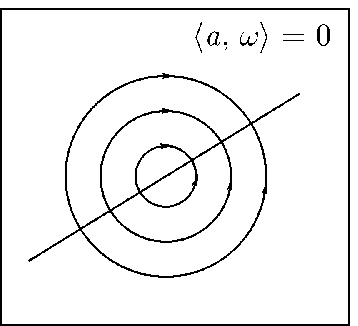}

G.\,K.\,Suslov has considered a particular case of the problem, when
the body is not under action of exterior forces:~$V\equiv 0$. In this case
the first equation of~(\ref{5.1}) is closed relatively to~$\om$. We can
show that it is integrable by quadratures (see [11], ch.~53). The
analysis of these quadratures shows that if~$\ba$ isn't an eigenvector
of the inertia operator, then all trajectories~$\om(t)$ approach
asymptotically as~$t\to\pm\fy$ to some fixed straight line on the
plane~$\<\ba,\,\om\>=0$ (see Fig.~1).
Consequently, the equation with respect to $\om$ and complete
system (\ref{5.1}) have
no invariant measure with continuous density. In this case theorem~1
isn't applicable, so, the question about the possibility to find the
vector~$\gam(t)$
by quadratures remains open. But if~${\bI\ba}=\mu\ba$ then
equations~(\ref{5.1}) have the additional integral:
the value of the kinetic moment is preserved
\eq*{
F_4=\<\bI\om,\,\bI\om\>.
}
Equations~(\ref{5.1}) are
integrable by theorem~1. However, this possibility may be easily
realized directly. It seems that in the most general case, the existence
of an invariant measure is connected with the hypothesis of proposition~5:
$\bI\ba=\mu\ba$. From now on, we suppose that this equality is fulfilled.

Now suppose that the body rotates in the homogeneous force field~$V=\<
{\bb},\,\gam\>$. If~$\<\ba,\,\bb\>=0$, then equations~(\ref{5.1}) have the
integral
\eq*{
F_4=\<\bI\om,\,\bb\>
}
consequently, they are integrable by quadratures. This case was indicated
by E.\,I.\,Kharlamova in her work~[12]. We are going to consider an
``opposite'' case, when~$\bb=\eps\ba$, $\eps\ne 0$. Without loss of generality
we can assume that the vector~$\ba$ has the components~$(0,\,0,\,1)$.
Taking into account the equation~$\om_3=0$, we obtain that two first
equations~(\ref{5.1}) have the following form
\eq*{
I_1\dot{\om}_1=\eps\gam_2, \q I_2\dot{\om}_2=-\eps\gam_1; \q
\om=(\om_1,\,\om_2,\,\om_3).
}
Therefore $I_1\ddot{\om}_1=\eps\dot{\gam}_2$,
$I_2\ddot{\om}_2=-\eps\dot{\gam}_1$.

Using the Poisson equations $\dot{\gam}_1=-\om_2\gam_3$, $\dot{\gam}_3=\om_1\gam_3$
we get
\begin{equation}
\label{5.2}
I_1\ddot{\om}_1=\eps\gam_3\om_1, \q I_2\ddot{\om}_2=\eps\gam_2\om_2.
\end{equation}
The energy integral
\eq*{
(I_1\om_1^2+I_2\om_2^2)\bigl/2+\eps\gam_3=h
}
makes it possible to express $\gam_3$ through $\om_1$ and~$\om_2$. After that,
equations~(\ref{5.2}) may be rewritten as the Lagrange equations
\begin{equation*}
\begin{gathered}
I_i^2\ddot{\om}_i=\pt{V}{\om_i}\Leftrightarrow\frac{d}{dt}\pt{L}{\dot{\om}_i}=
\pt{L}{\om_i} \q (i=1,\,2),\\
L=T_V, \q T=\frac{I_1^2\dot{\om}_1^2+I_2^2\dot{\om}_2}{2}, \q
V=\frac{1}{2}\Bigl(h-\frac{I_1\om_1^2+I_2\om_2^2}{2}\Bigr)^2.
\end{gathered}
\end{equation*}
These equations have the energy integral $T+V$. For
real motions its value is equal to~$\eps^2/2$. Let us emphasize that
unlike the reducing multiplier theory our reduction of equation~(\ref{5.1})
to Lagrange (or Hamilton) equations doesn't require the change of
time (cf.~[11]).

The change $I_i\om_i=k_i$ corresponding to the transition
from the angular velocity to the kinetic moment reduces the
considered problem of rotation of a body to the problem of motion
of a material point in the potential force field
\begin{equation}
\label{5.3}
\ddot{k}_i=-\pt{V}{k_i} \q (i=1,\,2), \q V=\frac{1}{2}\Bigl(h-
\frac{k_1^2 I_1^{-1}+k_2^2 I_2^{-1}}{2}\Bigr)^2.
\end{equation}
At $I_1=I_2$ we have the motion of point in a central field.
This motion corresponds to the well-known integrable ``Lagrange case''
of the generalized Suslov problem. As well as in Lagrange's
classical problem of a heavy symmetric top, the equations of motion
are integrable in this case in elliptical functions of time. If~$I_1\ne I_2$,
then the equations apparently have no additional analytical integral
independent of the energy integral. The following
observation confirms this supposition. Put formally~$I_1=-I_2=1$.
Then at~$h=0$ equations~(\ref{5.3}) practically coincide with the equations
of the Young-Mills homogeneous two-component model,
non-integrability of which is established in~[14].

If the value of $h$ is fixed, the point moves in the area defined
by the inequality~$V\le\eps^2/2$. For different~$h$, these areas
are shown in Fig~2.
The trajectories of vibrational motions, when one of the components~$k_1$
or~$k_2$ becomes zero, are especially interesting. These
motions are expressed through elliptical functions of time.

\wfig<bb=0 0 63.6mm 27.8mm>{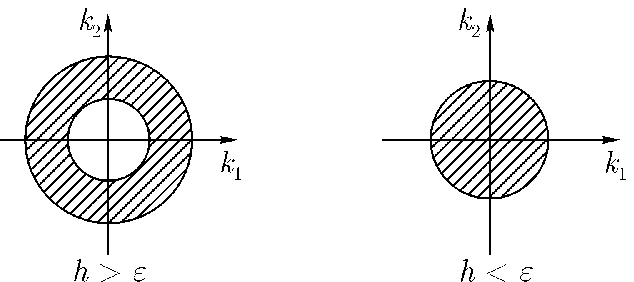}

One more case of integrability of (\ref{5.1}) is given by

\begin{teo}(3)
Suppose $\bI\ba$ and the potential $V(\gam)$ is defined by formula~\eqref{4.1}.
Then equations~\eqref{5.1} are integrable by quadratures.
\end{teo}

\proof
Let us show that equations (\ref{5.1}) have the Clebsch\f Tisseran
integral
\eq*{
F_4=\frac{1}{2}\<\bI\om,\,\bI\om\>-\frac{1}{2}\<A\gam,\,\gam\>, \q
A=\eps\bI^{-1}\det\bI.
}

Indeed,
\begin{multline*}
F_4=\<\bI\om,\,\gam\x\eps\bI\gam\>+\lm\<\ba,\,\bI\om\>+\<A\gam,\,\om\x\gam\>=\\
=\bigl\<\om,\,\bI(\gam\x\eps\bI\gam)\bigr\>+\lm\mu\<\ba,\,\om\>+\<\om,\,\gam\x
A\gam\>=\\
=\bigl\<\om,\,\bI(\gam\x\eps\bI\gam)+\gam\x A\gam\bigr\>=0,
\end{multline*}
since $\bI(\gam\x\bI\gam)=-(\gam\x\bI^{-1}\gam)\det\bI$.
To complete the proof, we have to take into account the conclusion
of proposition~5 and to use theorem~1.

Let us show, how one can explicitly integrate  equation (\ref{5.1}). For
definiteness, let~$\ba=(0,\,0,\,1)$ and~$\eps>0$, $I_3\ge I_1$, $I_3\ge
I_2$.
Then~(\ref{5.1}) may be presented as the following closed system of
four differential equations:
\begin{equation*}
\begin{gathered}
I_1\dot{\om}_1=\eps(I_1-I_2)\gam_2\gam_3, \q
I_2\dot{\om}_2=\eps(I_3-I_1)\gam_1\gam_3,\\
\dot{\gam}_1=-\om_2\gam_3, \q \dot{\gam}_2=\om_1\gam_3, \q
\gam_3^2=1-\gam_1^2-\gam_2^2.
\end{gathered}
\end{equation*}
Let us introduce the new time $\tau$ by formula $d\tau=\gam_3\,dt$ and
denote the
differentiating with respect to~$\tau$ by prime. Then equations of
motion take form of a linear system with constant coefficients
\eqa*{
I_1{\om'}_1 & =\eps(I_2-I_3)\gam_2, \q {\gam'}_2=\om_1,\\
I_2{\om'}_2 & =\eps(I_3-I_1)\gam_1, \q {\gam'}_1=-\om_2.
}
They can be presented in the equivalent form
\eqa*{
{} & {\gam''}_1+\lm_1^2\gam_1=0, \q {\gam''}_2+\lm^2_2\gam_2=0,\\
\lm_1^2 & =\eps(I_3-I_1)\bigl/I_2, \q \lm_2^2=(I_3-I_2)\bigl/I_1.
}
Put
\eq*{
\vfi_1=-\arctg\frac{\lm_1\gam_1}{\om_2}, \q
\vfi_2=\arctg\frac{\lm_2\gam_2}{\om_1}.
}
These variables are angle variables on two-dimensional invariant
tori with
\eq*{
{\vfi'}_1=\lm_1, \q {\vfi'}_1=\lm_2.
}
Consequently,
\eq*{
\dot{\vfi}_1=\lm_1/\Fi, \q \dot{\vfi}_2=\lm_2/\Fi; \q \Fi=(1-c_1^2\sin^2\vfi_1-
c_2^2\sin^2\vfi_2)^{-1/2}.
}
The constants $c_1$ and $c_2$ $(c_1^2+c_2^2\le 1)$
can be expressed as functions of constant values of the energy integral
and Clebsch\f Tisseran
integral. The remarkable property of this problem is the
fact that the ratio of frequencies~$\lm_1/\lm_2$ is independent of initial
data and depends only on the constants of parameters of the problem.
Consequently, if the number
\eq*{
\sqrt{\frac{(I_3-I_1)I_1}{(I_3-I_2)I_2}}
}
is rational, then all solutions are periodic; otherwise
practically all trajectories aren't closed (except degenerated
motions, when~$\gam_1\equiv 0$ or~$\gam_2\equiv 0$). Let~$\vfi_s(0)=a_s$. Then
\eq*{
t=\intl_0^\tau\frac{dx}{\sqrt{1-c_1^2\sin^2(\lm_1 x+a_1)-c_2^2\sin^2(\lm_2
x+a_2)}}.
}
If $c_1=0$ (or $c_2=0$) then $\gam_1$ and $\gam_2$ (and consequently,~$\om_1$,
$\om_2$, $\gam_3$)
are elliptical functions of time. This conclusion is true in the
case~$\lm_1=\lm_2$ (i.\,e., when~$I_1=I_2$ or~$I_3=I_1+I_2$) for all~$c_1$, $c_2$. In
the most general case the analytical character of the solutions is
essentially more complex. In conclusion, note that series~(\ref{2.5})
diverges in this problem if the irrational ratio~$\lm_1/\lm_2$ is
approximated by rational numbers anomalously fast.

\begin{rem*}
Equations (\ref{5.1}) are also integrable for potentials of
the general form
\eq*{
V(\gam)=\frac{1}{2}(c_{11}\gam_1^2+c_{22}\gam_2^2+c_{33}\gam_3^2+2c_{12}\gam_1\gam_2).
}
Using the change of time $d\tau=\gam_3\,dt$, the equations of motion are
reduced to the linear system
\eq*{
I_2\gam''_1=-\pt{V}{\gam_1}, \q I_1\gam''_2=-\pt{V}{\gam_2}; \q
\wt{V}=V\bigl|_{\gam_3^2=1-\gam_1^2-\gam_2^2}.
}
In the general case, potential~$V$ does not have a simple physical
interpretation.
\end{rem*}

\section{The first integrals used as constraints}


Let $L(\dot{x},\,x,\,t)$ be a Lagrangian of a nonholonomic system that
satisfies the following ``regularity'' condition: the quadratic form
\eq*{
\Bigl\<\frac{\d^2 L}{\d\dot{x}}\,\xi,\,\xi\Bigr\>
}
is positively definite. In particular, $\det\|L''_{\dot{x}\dot{x}}\|\ne 0$.
The constraints (not necessarily linear) are given by the equations
\begin{equation}
\label{6.1}
f_1(\dot{x},\,x,\,t)=\ldots=f_m(\dot{x},\,x,\,t)=0
\end{equation}
with independent co-vectors
\eq*{
\pt{f_1}{\dot{x}}\dts\pt{f_m}{\dot{x}}.
}
The equations of motion can be presented as the Lagrange equations
with the multipliers
\begin{equation}
\label{6.2}
\frac{d}{dt}\Bigl(\pt{L}{\dot{x}}\Bigr)-\pt{L}{x}=\sum_{s=1}^m\lm_s\pt{f_s}{\dot{x}},
\q f_1=\ldots=f_m=0.
\end{equation}

\begin{pro}(6)
If the functions $L,\,f_1\dts f_m$ satisfy the above
conditions, then a unique solution of~\eqref{6.2} corresponds to every
initial state that is permissible by constraints~\eqref{6.1}.
\end{pro}

Indeed, under these suppositions the multipliers $\lm_1\dts\lm_m$ are smooth
functions of~$\dot{x}$, $x$, $t$ by the explicit function theorem.

Now suppose equations (\ref{6.2}) have a first integral~$F(\dot{x},\,x,\,t)$.
We get the following

\begin{pro}(7)
If the co-vectors $\pt{F}{\dot{x}}$, $\pt{f_1}{\dot{x}}\dts\pt{f_m}{\dot{x}}$
are independent then~$x(t)$ is a solution of~\eqref{6.2} with the
constant value of the integral~$F=c$ if and only if this function is
a motion of mechanical system with the Lagrangian~$L$ and constraints
$f_1=\ldots=f_m=f_{m+1}=c$,
where~$f_{m+1}=F-c$.
\end{pro}

The sufficiency is obvious for $x(t)$ satisfies (\ref{6.2}) if we put~$\lm_{m+1}=0$.
On the contrary, let~$x(t)$ be a solution of a system of
form~(\ref{6.2}), where~$s$ ranges from~1 to~$m+1$. Let~$y(t)$ be
the unique motion of system~(\ref{6.2}) with the initial data
$y(0)=x(0)$, $\dot{y}(0)=\dot{x}(0)$.
Evidently,~$F\bigl|_{y(t)}=c$.
The function~$y(t)$ as well as~$x(t)$ satisfies the equations of
motion of the extended system with~$\lm_{m+1}=0$. To complete the proof,
it remains to use the conclusion of proposition~6.

Let us discuss one of possible applications of proposition 7. Suppose~$f_i$
are linear with respect to velocities and constraints~(\ref{6.1})
are non-integrable. If equations of motion have the linear integral~$F$,
then equations
\eq*{
f_1=\ldots=f_m=f_{m+1}=0 \q (f_{m+1}=F-c)
}
may turn out to be completely integrable. In this case the study of motions
that belong to the level set~$F=c$ is reduced to the investigation of
some holonomic system. We do not have to integrate here the
constraint equations, since the variables can be considered as redundant
one's, and the equations of motion may be written as Hamilton equations
in redundant variables (see~[11], [15]).

Let us consider as an example Suslov's problem in a homogeneous
force field in the Kharlamova integrable case. The
equations~$\<\ba,\,\om\>=0$ and~$\<\bI\om,\,\bb\>=0$ form an integrable field
of directions on the manifold of the rigid body positions (on the
group~SO(3)). Thus, Suslov's problem is reduced in this case
to a system with one degree of freedom. Though the one-dimensional
manifold of states of such system isn't closed in~SO(3) in the
general case.

If the constraints are non-linear with respect to velocities, it is
natural to use the energy integral
\eq*{
H(\dot{x},\,x)=\pt{L}{\dot{x}}\dot{x}-L
}
as the first integral\footnote{The equation of motion has the integral of energy,
if the constraints are homogeneous with respect to velocities and
the Lagrangian does not depend on time explicitly.}.

For example, let us consider the Appell\f Hamel system with the Lagrangian
\eq*{
L=\frac{1}{2}(\dot{x}^2+\dot{y}^2+\dot{z}^2)+gz, \q g=\const
}
and the non-linear constraint
\begin{equation}
\label{6.3}
\dot{x}^2+\dot{y}^2=k^2\dot{z}^2, \q k=\const\ne 0
\end{equation}
(see [16] and [17]). By the energy integral
\eq*{
\frac{1}{2}(\dot{x}^2+\dot{y}^2+\dot{z}^2)-gz=h
}
and (\ref{6.3}) we get the equation of the ``integrable'' constraint
\begin{equation}
\label{6.4}
\frac{\dot{z}(1+k^2)}{2}-gz=h.
\end{equation}
Consequently, the coordinate $z$ changes with the constant
acceleration~$g/(1+k^2)$.
Excluding non-linear integrable constraint~(\ref{6.4})
(i.\,e., considering~$z$ as a known function of time) we get a more
simple system with two degrees of freedom, the Lagrangian
\eq*{
\wt{L}=\frac{1}{2}(\dot{x}^2+\dot{y}^2)
}
and the constraint $\dot{x}^2+\dot{y}^2=f(t)$,
where~$f=k^2\dot{z}^2$ is a known quadratic function of time. The further
integration may be easily fulfilled.

\section{Symmetries of nonholonomic systems}


We suppose that the vector field ${\bv}(x)\ne 0$ is a symmetry field of
a nonholonomic system with Lagrangian~$L(\dot{x},\,x)$ and constraints
\eq*{
f_1(\dot{x},\,x)=\ldots=f_m(\dot{x},\,x)=0,
}
if the phase flow $g_v^s$ of the differential equation
\eq*{
\frac{dx}{dt}={\bv}(x)
}
preserves $L$ and $f_1\dts f_m$.

\begin{pro}(8)
A phase flow of a symmetry field converts
solutions of a nonholonomic system to solutions of the same system.
\end{pro}

\proof
By the theorem on rectification of trajectories, the phase
flow~$g_v^s$ in some local coordinates~$x_1\dts x_n$ is the following
one-parameter group
\eq*{
x_1\to x_1+s; \q x_2\to x_2\dts x_n\to x_n.
}
With respect to these variables, $L$ and $f_i$ do not depend on~$x_1$,
consequently, the equations of motion do not contain this variable, too.
This fact implies proposition~8.

In the case of integrable constraints, the symmetry field
corresponds to a linear with respect to velocities first integral
of the equations of motion. It is not so in the case of nonholonomic
systems.

\begin{pro}(9)
If $g_v^s$ preserves the Lagrangian and $v$
is the field of possible velocities, i.\,e.
\eq*{
\pt{f_1}{\dot{x}}{\bv}=\ldots=\pt{f_m}{\dot{x}}{\bv}=0,
}
then the equations of motion have the first integral
\eq*{
\pt{L}{\dot{x}}=\const.
}
This assertion {\rm(}``the Noether theorem''{\rm)} is discussed in~$[6]$, for example.
\end{pro}

\begin{teo}(4)
Suppose the equations of motion \eqref{6.2} have $n-m$ first
integrals~$f_{m+1}\dts f_n$.
If

$1)$ at points of the set $\t{E}_c=\{f_1=\ldots=f_m=0$, $f_{m+1}=c_{m+1}\dts f_n=c_n\}$
the Jacobian
\eq*{
\pt{(f_1\dts f_n)}{\dot{x}_1\dts\dot{x}_n},
}
is nonzero,

$2)$ there exist fields ${\bv}_1\dts{\bv}_{n-1}$ that are linearly independent
at all points $\t{E}_C$ and generate a solvable Lie algebra with
respect to the commutation operation, while their phase flows~$g^s_{v_i}$
preserve~$L$ and~$f_1\dts f_n$,

$3)$ there are no vectors $\dot{x}=\sum\lm_s{\bv}_s(x)$, $\lm_s\in\mR$
among solutions of the system of equations
\begin{equation}
\label{7.1}
f_1=\ldots=f_m=0, \q f_{m+1}=c_{m+1}\dts \q f_n=c_n,
\end{equation}
then solutions of \eqref{6.2} that belong to~$\t{E}_c$
are found by quadratures.
\end{teo}

\begin{rem*}
In some cases the existence of first integrals of nonholonomic
systems can be established by the following observation. Let~$F(\dot{x},\,x)$
be the first integral of a ``free'' holonomic system with Lagrangian~$L$.
This function is an integral of a nonholonomic system with the same
Lagrangian~$L$ and constraints~$f_1=\ldots=f_m=0$
in the case of
\eq*{
\Bigl(\frac{\d^2
L}{\d\dot{x}^2}\Bigr)^{-1}\pt{f_s}{\dot{x}}\cdot\pt{F}{\dot{x}}=0, \q 1\le
s\le m,
}
if $f_1=\ldots=f_m=0$.
This condition of invariancy is fulfilled for the Clebsch-Tisseran integral
in Suslov's problem (theorem~3). Besides, it is fulfilled for the
energy integral in the case of homogeneous constraints and for the
Noether integral~$\pt{L}{\dot{x}}\cdot{\bv}$, if the field $v$ is the field of
possible velocities (proposition~9).
\end{rem*}

\proof[of theorem $4$]
By the explicit function theorem we obtain
from~(\ref{7.1}) that
\begin{equation}
\label{7.2}
\dot{x}={\bv}_c(x).
\end{equation}
By conditions 2 and 3, the vectors ${\bv}_c,\,{\bv}_1\dts{\bv}_{n-1}$
are linearly independent at all points $x$. The phase flows~$g^s_{v_i}$
convert solutions of~(\ref{7.2}) to solutions of the same equation
(proposition 8).
To complete the proof, it remains to apply the well-known Lie theorem
on integrability by quadratures of systems of differential equations
(see, for example, [18]).

Let us consider as an illustrating example the problem of sliding
of a balanced skate on horizontal ice. One can choose units
of length, time and mass so that the Lagrangian would take the
following form:
\begin{equation}
\label{7.3}
L=\frac{1}{2}(\dot{x}^2+\dot{y}^2+\dot{z}^2).
\end{equation}
Here $x$, $y$ are the coordinates of the point of contact, $z$ is the
angle of rotation of the skate. The constraint equation is
\begin{equation}
\label{7.4}
f=\dot{x}\sin z-\dot{y}\cos z=0.
\end{equation}
The equations of motion have two first integrals
\begin{equation}
\label{7.5}
\dot{x}^2+\dot{y}^2+\dot{z}^2=h, \q \dot{z}=c.
\end{equation}
The second one is obtained by using proposition 9 with the help of
the vector field
${\bv}_3=(0,\,0,\,1)$. By~(\ref{7.4}) and~(\ref{7.5}) we obtain the field
${\bv}_{h,c}=\bigl(\sqrt{h-c^2}\,\cos z$, $\sqrt{h-c^2}\,\sin
z,\,c\bigr)$.
The fields~${\bv}_1=(1,\,0,\,0)$ and~${\bv}_2=(0,\,1,\,0)$
are commuting symmetry fields.
If~$c\ne 0$ then vectors~${\bv}_{h,c}$, ${\bv}_1$
and~${\bv}_2$ are linearly independent,
consequently, in this case we can apply theorem 4.
Let us emphasize that~${\bv}_1$ and~${\bv}_2$ do not generate the
conservation laws.

Theorem 4 impose strict restrictions on the nonholonomic system.
These restrictions can be weakened if we replace condition 2 by the
condition

2) for the fixed $c=(c_{m+1}\dts c_h)$
there exist $n-1$ linearly independent fields~${\bv}_i(x,\,c)$ that generate
a solvable Lie algebra and commute with~${\bv}_c(x)$.

Let us add to Lagrangian (\ref{7.3}) the term $-x/2$. Thus,
we have placed the skate onto an inclined plane. Equations~(\ref{7.4})--(\ref{7.5})
hold if we replace $x$ by $h+x$. Then the field~${\bv}_{h,c}$
is equal to
\eq*{
\bigl(\sqrt{h-c^2+x}\,\cos z, \; \sqrt{h-c^2+x}\,\sin z,\,c\bigr).
}
If $h$ and $c\ne 0$ are fixed, then the fields
\eq*{
{\bv}_1=\bigl(2\sqrt{h-c^2+x},\;-(\cos z)\bigl/c,\,0\bigr), \q {\bv}_2=(0,\,1,\,0)
}
and ${\bv}_{h,c}$ are independent, and all their commutators vanish.
In the same way one can solve the problem of rolling of a
homogeneous disk on a rough plane, the problem of rolling
of a ball in a vertical pipe and a series of other problems
of nonholonomic mechanics.

\section{Existence of an invariant measure}


The existence of an integral invariant with a positive density
is interesting not only from the standpoint of integration
of differential equations. It is interesting by itself, from
the standpoint of possible applications, for example, in ergodic theory.
We are going to consider the problem of existence of
an invariant measure for systems of differential equations.
We are especially interested in its applications to nonholonomic
mechanics.

By the theorem on rectification of trajectories, in a sufficiently
small neighborhood of an ordinary point  there always exists
an invariant measure with a smooth stationary density. Therefore,
the problem of existence of an invariant measure is especially interesting
near equilibriums as well as in sufficiently big domains
of the phase space, where trajectories have the property of
returning. Let us consider the first possibility.
Let the point $x=0$ be an equilibrium of an analytical system
of differential equations
\begin{equation}
\label{8.1}
\dot{x}=\Lm x+\ldots
\end{equation}

We say that a set of (complex) eigenvalues $\lm_1\dts \lm_n$ of the matrix
$\Lm$ is resonant, if~$\sum m_i\lm_i=0$
for some natural~$m_i$. Note that a weaker resonance
condition:~$\sum m_i\lm_i=0$
for some integer~$m_i\ge 0$ and~$\sum|m_i|\ne 0$
is usually used for investigation of
system~(\ref{8.1}) (for example, in the theory of
normal forms).

\begin{pro}(10)
If a set $\lm_1\dts\lm_n$ is not resonant then in a small
neighborhood of the point $x=0$ equation~\eqref{8.1} has no integral
invariant with an analytical density.

The non-resonance condition is fulfilled, for example, in the case
of $\Re\lm_i\ge 0$ $(\le 0)$ and ${\sum\Re\lm_i>0}$ $(<0)$.
\end{pro}

\proof
Let us expand the density $M(x)$ in a convergent series with
respect to homogeneous forms:
\eq*{
M=M_s+M_{s+1}+\ldots, \q s\ge 0.
}
Evidently, $M_s$ is the density of the integral invariant
for the linear system~$\dot{x}=\Lm x$. One can assume that~$\Lm$
as already reduced to the canonical Jordan form. Let us arrange the
monomials of the form~$M_s$ in some lexicographical order:
\eq*{
M_s=\sum_{m_i\ge 0 \atop m_1+\ldots+m_n=s}a_{m_1\ldots m_n}x_1^{m_1}\ldots
x_n^{m_n}.
}
It is evident that $\div M_s(\Lm x)$ is some form of the same degree.
By equating its coefficients to zero, we get a linear homogeneous system of
equations with respect to~$a_{m_1\ldots m_n}$. The determinant of this
system is equal to the product
\eq*{
\prod_{m_i\ge 0 \atop
m_1+\ldots+m_n=s}\bigl[(m_1+1)\lm_1+\ldots+(m_n+1)\lm_n\bigr].
}
This product is nonzero by supposition. Consequently, all
$a_{m_1\ldots m_n}=0$.\qed

\begin{rem*}
If a more strict condition of absence of resonant ratios
in the traditional sense is fulfilled then equation~(\ref{8.1}) has no first
integrals analytical in a neighborhood of the
point~$x=0$.
\end{rem*}

Let us consider as an example the problem on permanent rotations
of a convex rigid body with an analytical convex bound on a horizontal
absolutely rough plane (see~[4]). The motion of such body is described
by a system of six differential equations that have the integral of
energy and the geometrical integral. In a particular case, when
one of principal central axes of inertia of the body is
orthogonal to its surface, we have a one-parameter family of
stationary rotations about the vertical axis of inertia. Singular
points of equations of motion correspond to the stationary motions.
The characteristic equation has the following form:
\eq*{
\lm^2(a_4\lm^4+a_3\lm^3+a_2\lm^2+a_1\lm^1+a_0)=0.
}

The dependence of the coefficients $a_s$ on numerous parameters of
the problem is rather complicated; practically, they are arbitrary.
The existence of the double zero root
is connected with the existence of two independent integrals,
since the differentials of the energy integral and of the
geometrical integral are independent in the general case at points
that correspond to permanent rotations. Fixing the levels
of the first integrals we get differential equations on four-dimensional
manifolds that have no invariant measure with an
analytical density in the general case. Consequently, the initial
equations also have no invariant measure in a neighborhood
of stationary motions.

\wfig<bb=0 0 40.7mm 40.5mm>{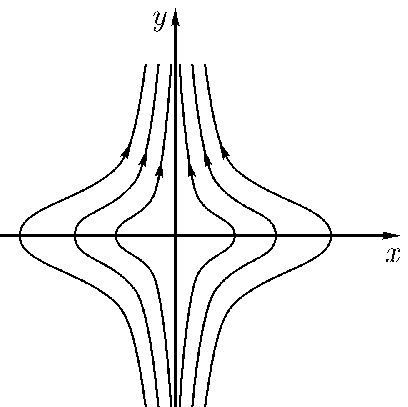}

Now let us consider the problem of existence of invariant measure for systems
of differential equations that are similar to integrable systems
that satisfy the conditions of theorem~1. It is natural to take
the constants of the first integrals~$I_1\dts I_{n-2}$ as
independent variables in a neighborhood of invariant tori
of a non-disturbed integral system and to take angle variables
$x,\,y\mod{2\pi}$ on the invariant tori. In these
variables, the perturbed system has the following form:
\begin{equation}
\label{8.2}
\begin{gathered}
\dot{I}_s=\eps f_s(I,\,x,\,y)+\ldots, \q s=1\dts n-2,\\
\dot{x}=\frac{\lm(I)}{\Fi(I,\,x,\,y)}+\eps X(I,\,x,\,y)+\ldots, \q
\dot{y}=\frac{\mu(I)}{\Fi(I,\,x,\,y)}+\eps Y(I,\,x,\,y)+\ldots
\end{gathered}
\end{equation}
We assume that all functions in the right-hand sides of these differential
equations are analytical in the direct
product~$D\x T^2$, where~$D$ is a domain in $\mR^{n-2}=\{I_1\dts
I_{n-2}\}$, $T^2=\{x,\,y\mod{2\pi}\}$; $\eps$
is a small parameter. For system~(\ref{8.2}), it is natural to consider the
problem of existence of an invariant measure, the density of which
is analytical with respect to $I$, $x$, $y$, $2\pi$\1periodic with
respect to $x$, $y$ and analytically depends on~$\eps$:
\begin{equation}
\label{8.3}
M=M_0+\eps M_1+\eps^2 M_2+\ldots
\end{equation}

The unperturbed problem has an invariant measure with the density $M_0$.
According to the well-known averaging principle, we average the right-hand
sides of~(\ref{8.2}) with respect to the measure $dm=\Fi\,dx\wedge dy$. As a result,
we get the closed system of equations for changing of slow variables~$I$
in the domain~$D$
\begin{equation}
\label{8.4}
\begin{gathered}
\dot{I}_s=\eps F_s(I), \q 1\le s\le n-2,\\
F_s=\frac{1}{\Lm}\intl_0^{2\pi}\intl_0^{2\pi}f_s\Fi\,dx\,dy, \q
\Lm=\intl_0^{2\pi}\intl_0^{2\pi}\Fi\,dx\,dy.
\end{gathered}
\end{equation}

\begin{pro}(11)
Suppose $m\lm(I)+n\mu(i)\not\equiv\, 0$
in domain~$D$ for all integer~$m$, $n$ that are not equal to zero
simultaneously. If averaged system~\eqref{8.4} has no invariant measure
with analytical density, then initial system~\eqref{8.2} also has no
invariant measure with density~\eqref{8.3}.

System \eqref{8.4} is simpler than \eqref{8.2}; the sufficient condition of
nonexistence of an invariant measure for~\eqref{8.4} is given by proposition~$10$.
\end{pro}

\proof[of proposition $11$]
Coefficients $M_0$ and $M_1$ of (\ref{8.3})
satisfy equations
\begin{equation}
\label{8.5}
\lm\pt{}{x}\frac{M_0}{\Fi}+\mu\pt{}{y}\frac{M_0}{\Fi}=0,
\end{equation}

\begin{equation}
\label{8.6}
\sum_s\pt{}{I_s}(M_0 f_s)+\pt {}{x} M_0 X+\pt{}{y}M_0 Y=-
\Bigl(\lm\pt{}{x}\frac{M_1}{\Fi}+\mu\pt{}{y}\frac{M_1}{\Fi}\Bigr).
\end{equation}
Since $\lm/\mu$ is irrational for almost all $I\in D$, equation~(\ref{8.5})
implies~$M_0=\Gam(I)\Fi$. Substituting this relation into~(\ref{8.6}) and
averaging with respect to $x$, $y$ we get the following equation:
\begin{equation}
\label{8.7}
\sum_s\pt{}{I_s}\Gam F_s=0.
\end{equation}
Consequently, $\Gam$ is the density of the integral invariant of (\ref{8.4}). It
remains to show that~$\Gam\not\equiv \,0$. If it is not true then~$M_0=0$.
But in this case the function~$M_1+\eps M_2+\ldots$ is the density
of an invariant measure for~(\ref{8.2}) If~$M_1\equiv 0$, this operation
may be repeated once more. The proposition is proved.

\begin{rem*}
One can show that (under conditions of proposition 11)
if averaged system~(\ref{8.4}) has no analytical first integral in $D$
then initial system~(\ref{8.2}) has no integral that can be expressed as
a series~$g_0+\eps g_1+\ldots$
with coefficients~$g_s$ analytical in~$D\x T^2$.
\end{rem*}

Let us consider in more details the particular case, when $n=3$. The
index $s$ may be omitted. Let~$F(I)\not\equiv \,0$. If~$F(I)=0$ at some
point of the interval $D$ then~(\ref{8.4}) evidently has no invariant
measure. Therefore, we assume that~$F(I)\ne 0$ in~$D$. Consider the following
Fourier expansions:
\eqa*{
\frac{X\Fi}{F} & =\sum X_{mn}(I)\exp i(mx+ny),\\
\frac{Y\Fi}{F} & =\sum Y_{mn}(I)\exp i(mx+ny),\\
\frac{f\Fi}{F} & =\sum f_{mn}(I)\exp i(mx+ny).
}
The resonant set $\Dl$ is the set of points $I\in D$, such that
\eq*{
\sum_{|m|+|n|\ne 0}\Bigl|\frac{a_{mn}}{m\lm+n\mu}\Bigr|^2=\fy, \q
a_{mn}=\frac{df_{mn}}{dI}+i(mX_{mn}+nY_{mn}).
}

\begin{pro}(12)
Suppose\nopagebreak

$1)$ $\lm(I)\bigl/\mu(I)\not\equiv\,\const$,\nopagebreak

$2)$ the intersection $\Dl\cap D$ is not empty.\nopagebreak

Then \eqref{8.2} has no integral invariant with density \eqref{8.3}.
\end{pro}

Indeed, correlation (\ref{8.7}) implies $\Gam=c/F$, where $c=\const$.
Let
\eq*{
\frac{M_1}{\Fi}=\Sg b_{mn}(I)\exp i(mx+ny).
}
Equation (\ref{8.6}) gives us the set of correlations
\eq*{
-(m\lm+n\mu)b_{mn}=ca_{mn}.
}
Let $I\in\Dl$. Then the condition
\eq*{
\Sg\,|b_{mn}|^2<\fy
}
implies $c=0$. \qed

The author is grateful to professor V.\,F.\,Zhuravlev who has read
the paper and made a number of remarks.

\end{document}